\documentstyle[12pt]{article}  
  
\title{ UNIVERSAL QUANTIFICATION FOR SELF-ORGANIZED CRITICALITY IN ATMOSPHERIC 
FLOWS   
}  
  
\author{A.Mary Selvam \\  
Indian Institute of Tropical Meteorology ,Pune 411 008, India \\
\\Proc. Conf. Patterns,Nonlinear Dynamics and\\
 Stochastic Behaviour in Spatially Extended Complex Systems,\\
 October 24-28,1997,Budapest,Hungary}
\date{     }  
\begin{document}  
  
\maketitle  
\begin{abstract}  
Atmospheric flows exhibit selfsimilar fluctuations on all scales(space-time)
ranging from climate(kilometers/years) to turbulence(millimeters/seconds)
manifested as fractal geometry to the global cloud cover pattern concomitant
with inverse power law form for power spectra of temporal fluctuations.
Selfsimilar fluctuations implying long-rang correlations are ubiquitous
to dynamical systems in nature and are identified as signatures of
self-organized criticality in atmospheric flows. Also, mathematical models
for simulation and prediction of atmospheric flows are nonlinear and
computer realization give unrealistic solutions because of deterministic
chaos, a direct consequence of finite precision round-off error doubling
for each iteration of iterative computations incorporated in long-term
numerical integration schemes used for model solutions An alternative
non-deterministic cell dynamical system model [1] predicts,
(a): the observed self organized criticality as a consequence of quantumlike
mechanics governing flow dynamics,.(b):atmospheric flows trace an overall
logarithmic spiral trajectory with the quasiperiodic Penrose tiling pattern
for the internal structure,(c): eddy circulation structure follows Kepler's
third law of planetary motion and results in inverse law form for
centripetal acceleration. The inertial masses representing the eddy
circulation therefore follow laws analogous to the Newton's inverse
square law for gravitation. The model is similar to a superstring model
for subatomic dynamics[2] which unifies quantum mechanical and
classical concept and incorporates gravitational forces along with
nuclear and electromagnetic forces.
\begin{verbatim}
[1]Mary Selvam et.al.,Int'l. J.climatol. 16(1996)1.

[2]Kaku, New Scientist 18 Jan. 1997,32  
\end{verbatim}  
\end{abstract}  
  
\end{document}